 \documentclass[12pt, draftclsnofoot, onecolumn,comsoc]{IEEEtran}

\usepackage{amsmath, amsfonts, amssymb, graphicx}
\usepackage{subfigure}
\usepackage{multirow}
\usepackage{url}
\usepackage{color}
\usepackage{algorithm, algorithmicx, algpseudocode}

\newtheorem{thm}{Theorem}

\newtheorem{lem}[thm]{Lemma}

\DeclareMathOperator{\E}{E}

\newcommand{\bs}{\boldsymbol}
\newcommand{\mbf}{\mathbf}
\DeclareMathOperator{\st}{subject\ to}

\begin{document}
\title{Distributed power allocation for D2D communications underlaying/overlaying OFDMA cellular networks}

\author{Andrea Abrardo, Marco Moretti, \textit{Member, IEEE}}
\maketitle

\begin{abstract}
The implementation of device-to-device (D2D) underlaying or overlaying pre-existing cellular networks has received much attention due to the potential of enhancing the total cell throughput, reducing power consumption  and increasing the instantaneous data rate.
In this paper we propose a distributed power allocation scheme for D2D OFDMA communications and, in particular, we consider the two operating modes amenable to a distributed implementation: dedicated and reuse modes. The proposed schemes address the problem of maximizing the users' sum rate subject to power constraints, which is known to be nonconvex and, as such, extremely difficult to be solved exactly. We propose here a fresh approach to this well-known problem, capitalizing on  the fact that the power  allocation problem can be modeled as a potential game. Exploiting the potential games property of converging under better response dynamics, we propose two fully distributed iterative algorithms, one for each operation mode considered, where each user updates sequentially and autonomously its power allocation.
Numerical results, computed for several different user scenarios, show that the proposed methods, which converge to one of the local maxima of the objective function, exhibit performance close to the maximum achievable optimum and outperform other schemes presented in the literature.
\end{abstract}

\section{Introduction}
The exponential growth of mobile radio communications has lead to a pressing demand for higher data rate of wireless systems and, more generally, it has brought up the necessity of improving the whole network performance. Accordingly, a large part of recent efforts of the research community has been focused on  increasing the spectral efficiency of wireless systems. This can be attained in several different ways exploiting in a way or the other the inherent diversity of mobile communications: for example by using a large number of antennas, or by exploiting the knowledge of the propagation channel at the transmitter to best adapt the usage of radio resources or by optimally sharing the existing spectrum with a larger number of users. At the same time, the development of advanced and spectrally efficient communication techniques has called for the deployment  of more effective interference management schemes with a great emphasis on network densification techniques \cite{Bhushan2014} as in the Third Generation Partnership Project (3GPP) Long Term Evolution Advanced (LTE-Advanced) systems \cite{LTE-A1}.
\par
In this scenario, the concept of device-to-device (D2D) communication underlaying or overlaying pre-existing cellular networks has received much attention due to the potential of enhancing the total cell throughput, reducing power consumption  and increasing the instantaneous data rate \cite{Enable-D2D}-\nocite{WenAndZhu2013}\cite{Guo:15}. Thanks to the D2D paradigm, user equipments (UEs) are able to communicate with each other over direct links, so that new proximity-based services,  content caching is just one example  \cite{Ming2016}, can be  implemented and evolved nodeB (eNB)  base stations can offload part of the traffic burden  \cite{D2D-discovery,Dynamic-PC}.
\par
In particular, the implementation of D2D communications in coexistence with cellular networks presents also several challenges and how to share the existing spectrum is one of the most important.
In order to provide the system with  maximum flexibility, D2D communications should be able to operate in the following multiple modes \cite{Survey}: (\emph{i})  \emph{Dedicated} or \emph{overlay} mode, when the cellular network allocates a fraction of the available resources for the exclusive use of D2D devices; (\emph{ii}) \emph{Reuse} or \emph{underlay} mode, when D2D devices use some of the radio resources together with the UEs of the cellular network; (\emph{iii}) \emph{Cellular mode} when D2D traffic passes though the eNB, as in traditional cellular communications. Among these,  reuse mode is potentially the best in terms of spectral efficiency, since it allows more than one user  to communicate over the same channel within each cell. In this case, mitigation of the  interference between cellular and D2D communications is a critical issue: good interference management algorithms can increase the system capacity, whereas poor interference management  may have catastrophic effects on the system performance \cite{Survey}.
In details, the authors of \cite{D2D_underlay1}-\nocite{Power-control1}\cite{Power-control2} have shown that by using proper power allocation strategies, the sum rate for orthogonal and non-orthogonal resource sharing modes can be significantly improved.
\par
All the above-mentioned existing resource allocation literature for D2D communications refers to scenarios where the eNB is responsible for managing the radio resources of both cellular and D2D users. Nevertheless, in most recent systems like LTE-A, this centralized approach might not be viable. In facts, the channels of a macro cell may be reused by user-deployed nodes such as  home NodeBs,  femto base stations or D2D communication nodes and due to the potentially large number of nodes within a cell, the growing complexity of the schedulers in existing nodes and the user requirements on plug-and-play deployment, there is a growing need for distributed radio resources allocation algorithms. On the other hand, distributed techniques, which tend to be implemented iteratively, have often the drawback of achieving sub-optimal results and exhibiting slow convergence \cite{Moretti2014}. In this case a key role is played by the amount of information that the various node share with each other.

\subsection{ Paper Contributions and Outline}
In this paper we consider a distributed power allocation scheme for D2D OFDMA communications and in particular we consider the two operating modes amenable to a distributed implementation: dedicated and reuse modes. 
The proposed schemes address the problem of maximizing the users' sum rate subject to power constraints. Although the literature is rich of contributions for this particular type of problems \cite{YuRhee2004}\nocite{Papandriopoulos2009}-\cite{Javan2014}, the problem is known to be nonconvex and, as such, extremely difficult to be solved exactly \cite{Power-control6}, \cite{Zhang2014}. In order to simplify the NP-hard resource allocation problem, most of the existing works assume that at most one D2D user can access the channel or, similarly, that resource reuse is not permitted among D2D terminals \cite{Yu:11}\nocite{Zulhasnine:10,Feng2013,Phunchongharn2013}-\cite{Guanding2014}. However, such a simplification yields poor spectral efficiency when more than one D2D terminal is admitted in the cell, e.g., in dense cellular networks. Recent works have started to consider sub-optimal centralized approaches where more than one D2D can share the same channel, e.g., \cite{Zhang2014}, \cite{C.Xu2013}.

We pursue here a fresh approach which leads to a distributed implementation best suited for D2D communications. In detail, these are the main contributions of this paper:
\begin{itemize}
\item We show that the rate maximization power allocation problem can be formulated as a potential game. Exploiting the potential games property of converging under better response dynamics, we propose two fully distributed iterative algorithms, one for each operation mode considered, where each user computes sequentially and autonomously its power allocation;
\item We prove the convergence of the distributed problem for the D2D \emph{dedicated} mode to a local maximum of the sum rate. By  linearizing the log function with its first order Taylor expansion, each user's  objective function is split in two terms: a logarithmic one that accounts for the users' own throughput and a linear one that can be interpreted as the penalty cost for using a certain resource, due to the interference generated for the other users. Such costs are evaluated as proposed in \cite{MP1}, where each eNB is a player of a non-cooperative game, and the payoff function is the total cell throughput;
\item For D2D \emph{reuse} mode the allocation problem is formulated with an additional requirement for each subcarrier so that the total interference generated at the base station by the D2D nodes does not exceed a given threshold. Accordingly, after finding the optimal solution, which is too complex for practical implementation,  we propose a heuristic algorithm, which builds on the power allocation algorithm devised for the dedicated D2D mode to find a feasible solution.

\item We discuss about possible practical implementation of the proposed allocation schemes. In particular, we first propose an approach based on the  exchange of messages between D2D terminals and the eNB. Each message carries the cost needed to evaluate the negative impact, in terms of global utility, of using a resource with a given power. In alternative, in order to avoid the protocol overhead resulting from network-wide message passing, we propose  a second approach based on the use of a broadcast sounding signal. In this case, the required information to perform power allocation can be gathered from interference measurements, without requiring neither message passing nor additional channel gain estimations.
\end{itemize}
Numerical simulations, carried out for several different user scenarios, show that the proposed methods, which converge to one of the local maxima of the objective function, exhibit performance close to the maximum achievable optimum and outperform other schemes presented in the literature. Moreover, comparisons between the proposed allocation schemes in a typical cellular scenario show the superiority of the reuse mode, thus proving the effectiveness of the proposed allocation schemes in exploiting the available radio resources.

The remainder of this paper is organized as follows. Section II sets the background by introducing the D2D paradigm and signal model. Sections III describes the power allocation algorithm for the dedicated mode and presents its solution based on a game theoretic approach.  Section IV addresses the problem of power allocation for the reuse mode. In Section V we discuss some implementation aspects of the proposed algorithms. Numerical results are illustrated in Section VI and conclusions are drawn in Section VII.
\section{Background}
\label{Sec:Problem}
\begin{figure}[ht]
\begin{center}
\includegraphics[width=0.8\hsize]{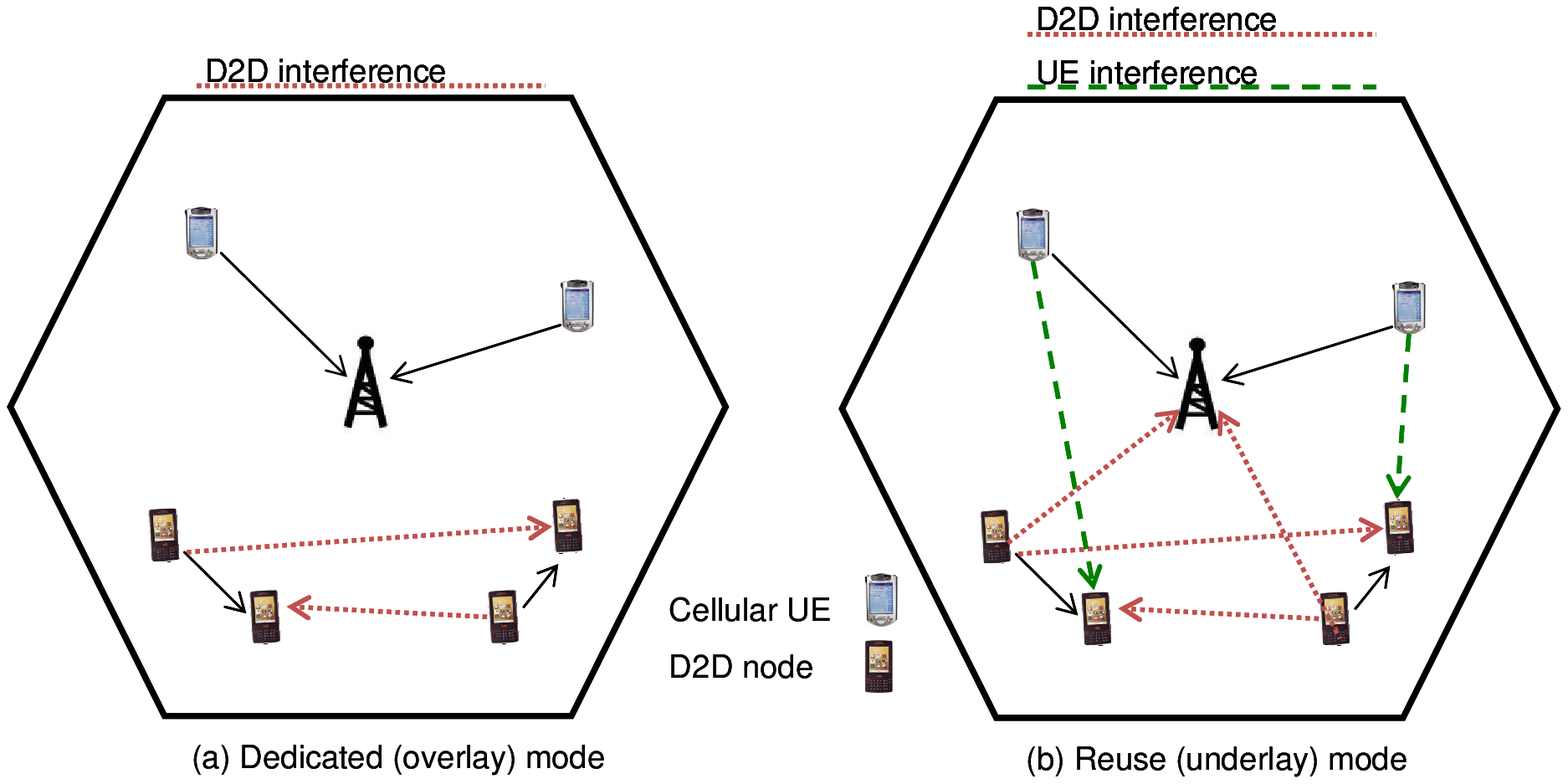}
\caption{
D2D communication modes: (a) \emph{Dedicated} mode, where D2D connections are assigned a fraction of the total available bandwidth, so that there is no  interference between cellular and D2D terminals; (b) \emph{Reuse} mode, where the whole uplink bandwidth is available to each D2D terminal, so that D2D nodes and UE terminals interfere with each other.
}
\label{F0}
\end{center}
\end{figure}

The two D2D scenarios considered in this paper, namely, dedicated and reuse D2D transmission modes, are illustrated in Fig. \ref{F0}. In particular, we envisage a cellular scenario where cellular and D2D connections coexist in the same cell and transmit over the same bandwidth. In  dedicated mode  a fraction of the total available bandwidth is assigned exclusively to  D2D  transmissions, so that interference between cellular and D2D terminals is completely avoided. The  interference between D2D connections is managed through distributed power allocation among D2D terminals, without requiring any cellular network control. Nevertheless, in order to avoid interference with cellular terminals located in adjacent cells, we assume that a  power mask, i.e., a maximum transmitting power, is imposed to each D2D terminal on each subcarrier.

In reuse mode  the whole uplink bandwidth is available to each D2D terminal so that D2D nodes and UEs are free to interfere with each other. Hence, in this mode the interference from D2D communications to the cellular receivers at the eNB must be controlled to prevent it from disrupting the QoS of cellular communications. This requires a form of centralized control, which actively involves the cellular network. Conversely, interference from cellular terminals to the D2D receivers is treated as uncontrollable additional noise.

Although the algorithms we propose are distributed and operate autonomously at the D2D nodes, it is the eNB \cite{Survey}  that is in charge of choosing  which of the two D2D modes is selected on the base of several factors such as available radio resources, network congestion and number of active mobile terminal.

To elaborate, we then consider to have a set $\mathcal{K}=\left\{1,\dots,K\right\}$ of $K$  D2D terminals that transmit over the set  $\mathcal{N}=\left\{1,\dots,N\right\}$ of  shared OFDMA channels. For each D2D terminal that transmits  there is another one that receives to  form a D2D couple, so that $H_{k,i}^{n}$ is the complex channel gain on subcarrier $n$ between the transmit node of the D2D couple $k$ and the receive node of  the D2D couple $i$. Assuming perfect synchronization, the signal at the $k$-th receiver on link $n$ is
\begin{equation}
Y_{k,n} = H_{k,k}^{n}S_{k,n}+ \sum\limits_{j\in \mathcal{K}\setminus k}H_{j,k}^{n}S_{j,n}+W_{k,n},
\end{equation}
where $S_{k,n}$ is the transmitted symbol, which we assume to be a zero-mean Gaussian distributed variable with power $p_{k,n}=\E\left\{|S_{k,n}|^{2}\right\}$  and $W_{k,n}$ is an additive zero-mean Gaussian disturb with variance $\sigma^{2}_{k,n}$, which, to a first approximation, includes the thermal noise and the interference from the infrastructured network.
Accordingly, employing the Shannon capacity formula, the throughput of the $k$-th couple over the $N$ available links is
\begin{equation}
R_{k}(\mathbf{p}_{k})=\sum\limits_{n\in\mathcal{N}}\log_{2}\left(1+\frac{G_{k,k}^{n}p_{k,n}}{\sum\limits_{j\in \mathcal{K}\setminus k}G_{j,k}^{n}p_{j,n}+\sigma_{k,n}^{2}}\right),
\label{eq:rateUserk}
\end{equation}
where $G_{k,k}^{n}=|H_{k,k}^{n}|^{2}$, $\mathbf{p}_{k}=[p_{k,1},p_{k,2},\dots,p_{k,N}]\in  \mathcal{P}_{k}$ is the vector stacking the power transmitted on the $N$ subcarriers by user $k$. The set $\mathcal{P}_{k}=\{\mathbf{p}_{k}\in[0,P_{k,1}]\times  [0,P_{k,2}] \times \dots \times [0,P_{k,N}]\}$ is the set of admissible power levels for user $k$ where $P_{k,n}$ is the power mask for user $k$ on subcarrier $n$.

\section{D2D Dedicated Mode: Rate Maximization Under a Power Constraint}
In  dedicated operation mode, the D2D nodes transmit over a fraction of the bandwidth which is dedicated exclusively to D2D transmissions. In this case we focus on the problem of finding the power allocation  that maximizes the sum of the rates of the D2D network with a power constraint per user.  Since we are considering a distributed  scenario, where each device tries to optimize its performance with a strategy that is influenced by other users' decisions, the presence of interference  greatly complicates the  problem with respect to the standard waterfilling solution.

\subsection{Joint Optimal Problem}
The problem of  jointly maximizing the overall rate of the D2D network (joint rate maximization problem, JRMP) can be formulated as
\begin{align} \label{eq:rateMaxJoint}
R(\mathbf{p}^{\ast})=\max\limits_{\mathbf{p}\in\mathcal{P}} &\sum \limits_{k \in\mathcal{K}} \sum\limits_{n \in \mathcal{N}}\log_{2}\left(1+\frac{G_{k,k}^{n}p_{k,n}}{\sum\limits_{j\in \mathcal{K}\setminus k}G_{j,k}^{n}p_{j,n}+\sigma_{k,n}^{2}}\right) \nonumber \\
&\st\\
&\sum_{n\in\mathcal{N}}p_{k,n}\leq P_{k} \quad k\in\mathcal{K} \nonumber
\end{align}
where $P_{k}$ is the maximum power constraint for the $k$th D2D node, $\mathbf{p}=[\mathbf{p}_{1},\mathbf{p}_{2},\dots,\mathbf{p}_{K}]$  and $\mathcal{P}=\mathcal{P}_{1}\times\mathcal{P}_{2}\times\dots\times\mathcal{P}_{K}$.
\par
JRMP is a well-studied allocation problem, which, as clearly shown in \cite{Papandriopoulos2009} in another framework,  is not convex and therefore standard solvers can not be directly applied to investigate its solution. In particular, considering that we are dealing with a D2D scenario, where distributed independent devices communicate with each other, we need to implement a distributed solution and the performance of a centralized allocator would only represent a performance bound rather than a viable practical option.
Therefore, instead of trying to solve the joint centralized problem employing the classical tools of convex optimization, we  invoke some important game theoretic  results about \emph{potential games} to find a solution for  the rate maximization in \eqref{eq:rateMaxJoint} and we propose a distributed iterative solution, that indeed requires the exchange of some information between the nodes, but that can be implemented locally at each transmitter.

As for any iterative strategy, two are the major concerns: the optimality of the algorithm and its convergence. Regarding the optimality, since the original problem is not convex, it might admit the existence of several local maxima and the same applies to the iterative solution, which is not guaranteed to converge to the global optimum. Algorithm's convergence is a major issue for this type of problems. For instance,  \emph{iterative waterfilling} \cite{YuRhee2004}, i.e., the distributed approach where each user aims selfishly at maximizing its own rate  individually, is known to converge only when interference does not exceed a certain critical level. In the following we prove that the distributed solution always converges to a local maximizer of the objective function in \eqref{eq:rateMaxJoint} regardless of the level of interference.

\subsection{Game theoretic formulation}
First of all we need few definitions. A game $\mathcal{G}(\mathcal{K},\{\mathcal{S}_{k}\}_{k\in\mathcal{K}},\{U_{k}\}_{k\in\mathcal{K}})$ is described by the set of players $\mathcal{K}$, the set $\mathcal{S}_{k}$ of all possible strategies and the utility function $U_{k}$ for each player $k\in\mathcal{K}$. Moreover, a set of strategies $s_{1}^{\ast},s_{2}^{\ast},\dots,s_{K}^{\ast}$ is a Nash equilibrium (NE), if  no user has any benefit to change individually its strategy, i.e.
\begin{equation}
U_{k}(s_{k}^{\ast},\mathbf{s}_{-k}^{\ast})>U_{k}(x_{k},\mathbf{s}_{-k}^{\ast}) \quad \forall x_{k}\ne s_{k}^{\ast},\forall k\in\mathcal{K}
\end{equation}
where $x_{k}\in\mathcal{S}_{k}$ is an arbitrary strategy of player $k$ and $\mathbf{s}_{-k}^{\ast}$ are the joint strategies of the other $K-1$ players.
The \emph{best response} dynamics of player $k$ are the set of strategies which maximize the payoff of player $k$ given its opponents strategies $\mathbf{s}_{-k}$. \emph{Better response} dynamics  for player $k$ employing strategy $y_{k}$ are the set of strategies $x_{k}\in\mathcal{S}_{k}$ such that $U_{k}(x_{k},\mathbf{s}_{-k})  > U_{k}(y_{k},\mathbf{s}_{-k})$.
\par
One particular  class of games is represented by \emph{potential} games, which are games in which the preferences of all players are aligned with a global objective. A game $\mathcal{G}(\mathcal{K},\{\mathcal{S}_{k}\}_{k\in\mathcal{K}},\{U_{k}\}_{k\in\mathcal{K}})$ is an \emph{exact} potential game  if it exists a potential function $f:\mathcal{S}_{1}\times\mathcal{S}_{2}\times\dots\times\mathcal{S}_{K}\mapsto\mathbb{R}$ such that for any two arbitrary strategies $x_{k},y_{k}\in\mathcal{S}_{k}$ the following equality holds
\begin{equation}
\label{eq:potentialFunction}
U_{k}(x_{k},\mathbf{s}_{-k})-U_{k}(y_{k},\mathbf{s}_{-k})=f(x_{k},\mathbf{s}_{-k})-f(y_{k},\mathbf{s}_{-k}) \quad \forall k \in\mathcal{K}.
\end{equation}
In a  potential game, where strategy sets are continuous and compact and the game is played sequentially, best/better response dynamics always converge from any arbitrary initial outcome to a NE, which is also a maximizer of the potential function \cite{Monderer1996}.
\par
\begin{thm} The power allocation game $\mathcal{G}\big(\mathcal{K},\{\tilde{\mathcal{P}}_{k}\},R(\mathbf{p}_{k},\mathbf{p}_{-k})\big)$ is an exact potential game.
\end{thm}
\begin{IEEEproof}
Let us consider the game $\mathcal{G}\big(\mathcal{K},\{\tilde{\mathcal{P}}_{k}\},R(\mathbf{p}_{k},\mathbf{p}_{-k})\big)$, where the players are the $K$ D2D rx-tx couples, the set of strategies for player $k$ is  $\tilde{\mathcal{P}}_{k}=\left\{\mathbf{p}_{k}\in\mathcal{P}_{k}\vert \sum_{n\in\mathcal{N}}p_{k,n}\leq P_{k}\right\} $, the set of all possible power profiles that meet the power constraint $P_{k}$, and the payoff function is $R(\mathbf{p}_{k},\mathbf{p}_{-k})$, where the power vector profile $\mathbf{p}_{-k}$, as customary in the game-theoretic literature, denotes the vector of the powers of all users but the $k$th one.
Since  the rate of the whole system $R(\mathbf{p}_{k},\mathbf{p}_{-k})$ is the payoff for each player $k\in\mathcal{K}$, the utility function is the same for all players. As a result it follows that $R(\mathbf{p})$ satisfies \eqref{eq:potentialFunction} and as such is a potential function of the game $\mathcal{G}\big(\mathcal{K},\{\tilde{\mathcal{P}}_{k}\},R(\mathbf{p}_{k},\mathbf{p}_{-k})\big)$.
\end{IEEEproof}
\par
The best response dynamic for user $k$ in the game $\mathcal{G}\big(\mathcal{K},\{\tilde{\mathcal{P}}_{k}\},R(\mathbf{p}_{k},\mathbf{p}_{-k})\big)$ is  the solution of the following distributed rate maximization problem (DRMP)
\begin{align}
\mathbf{p}_{k}^{\ast}=&\arg\max\limits_{\mathbf{p}_{k}\in\mathcal{P}_{k}}\underbrace{ \sum\limits_{n \in \mathcal{N}}\log_{2}\left(1+\frac{p_{k,n}}{i_{k,n}}\right)}_{\text{a)}}\nonumber\\
&+\underbrace{\sum\limits_{\ell \in \mathcal{K}\setminus k}\sum\limits_{n \in \mathcal{N}}\log_{2}\left(1+\frac{p_{\ell,n}}{i_{\ell,k,n}+\frac{G_{k,\ell}^{n}}{G_{\ell,\ell}^{n}}p_{k,n}}\right)}_{\text{b)}} \label{eq:rateMaxDist}\\
& \st \nonumber\\
& \sum_{n\in\mathcal{N}}p_{k,n}\leq P_{k} \quad k\in\mathcal{K} \nonumber
\end{align}
where the interference terms are  $i_{k,n}=\frac{\sum\limits_{j\in \mathcal{K}\setminus k}G_{j,k}^{n}p_{j,n}+\sigma_{k,n}^{2}}{G_{k,k}^{n}}$ and $i_{\ell,k,n}=\frac{\sum\limits_{j\in \mathcal{K}\setminus \{k,\ell\}}G_{j,\ell}^{n}p_{j,n}+\sigma_{\ell,n}^{2}}{G_{\ell,\ell}^{n}}=i_{\ell,n}-\frac{G_{k,\ell}^{n}}{G_{\ell,\ell}^{n}}p_{k,n}$.
\par
The objective function  \eqref{eq:rateMaxDist} is formulated to underline the fact that allocating power to user $k$ has two separate effects: a) it contributes to the rate for user $k$ and b)  it also affects the way that $k$ interferes with all other users. It is important to notice that, although the objective function of JRMP and of the $k$th DRMP are \emph{exactly} the same, for the latter  problem the optimization is performed only with respect to the power of the $k$th user and not jointly for all users as in \eqref{eq:rateMaxJoint}.
Unfortunately, the optimization in \eqref{eq:rateMaxDist} is still not convex and can not be easily solved. Accordingly, we replace the part b) of  \eqref{eq:rateMaxDist} with its first order Taylor expansion  $f(\mathbf{x})\approx f(\mathbf{x}_{0})+\nabla f^{T}(\mathbf{x}_{0})(\mathbf{x}-\mathbf{x}_{0})$, so that the  objective function of the $k$th DRMP  can be approximated around the vector $\mathbf{p}_{k}(0)$ by the function $\tilde{R}(\mathbf{p}_{k},\mathbf{p}_{-k};\mathbf{p}_{k}(0))$ as
\begin{align}\label{eq:rateMaxDistApprox}
R(\mathbf{p}_{k},\mathbf{p}_{-k})\approx&\tilde{R}(\mathbf{p}_{k},\mathbf{p}_{-k};\mathbf{p}_{k}(0))=\sum\limits_{n \in \mathcal{N}}\log_{2}\left(1+\frac{p_{k,n}}{i_{k,n}}\right)\nonumber\\
&+\sum\limits_{\ell \in \mathcal{K}\setminus k}\sum\limits_{n \in \mathcal{N}}\log_{2}\left(1+\frac{p_{\ell,n}}{i_{\ell,k,n}+\frac{G_{k,\ell}^{n}}{G_{\ell,\ell}^{n}}p_{k,n}(0)}\right)+\sum\limits_{n \in \mathcal{N}}\alpha_{k,n}\Big(p_{k,n}-p_{k,n}(0)\Big)
\end{align}
where
\begin{align}
\label{eq:Gradient}
\alpha_{k,n}=& \sum\limits_{\ell \in \mathcal{K}\setminus k}\frac{\partial }{\partial {p}_{k,n}}\log_{2}\left.\left(1+\frac{p_{\ell,n}}{i_{\ell,k,n}+\frac{G_{k,\ell}^{n}}{G_{\ell,\ell}^{n}}p_{k,n}}\right)\right\vert_{p_{k,n}=p_{k,n}(0)}\nonumber \\
=&- \sum\limits_{\ell \in \mathcal{K}\setminus k}\frac{G_{\ell,\ell}^{n}G_{k,\ell}^{n}p_{\ell,n}}{\ln2\left(G_{\ell,\ell}^{n}i_{\ell,k,n}+G_{k,\ell}^{n}p_{k,n}(0)\right)\left(G_{\ell,\ell}^{n}i_{\ell,k,n}+G_{\ell,\ell}^{n}p_{\ell,n}+G_{k,\ell}^{n}p_{k,n}(0)\right)}
\end{align}
Intuitively, the term $\alpha_{k,n}$, which is the $n$th element of the gradient of $R(\mathbf{p}_{k},\mathbf{p}_{-k})$ computed in $\mathbf{p}_{k}(0)$, represents the sensitivity of all other users  to the variations of the power of user $k$: by construction $\alpha_{k,n}$ is always negative and any increment of $p_{k,n}$ increases the rate of user $k$ on subcarrier $n$ but is coming with the negative penalty $\alpha_{k,n}p_{k,n}$.
The rate approximation in \eqref{eq:rateMaxDistApprox} is the sum of a convex function and an affine function in $\mathbf{p}_{k}$ and therefore is a convex function. Neglecting the terms not dependent on $\mathbf{p}_{k}$ and thus irrelevant to the optimization, the approximated DRMP (ADRMP) optimization can be written as
\begin{align}
\max\limits_{\mathbf{p}_{k}\in\mathcal{P}_{k}} & \sum\limits_{n \in \mathcal{N}}\log_{2}\left(1+\frac{p_{k,n}}{i_{k,n}}\right)+\sum\limits_{n \in \mathcal{N}}\alpha_{k,n}p_{k,n} \nonumber \\
& \st \label{eq:linProb}\\
& \sum_{n\in\mathcal{N}}p_{k,n}\leq P_{k} \nonumber
\end{align}
Appendix A illustrates the procedure to compute the  solution of \eqref{eq:linProb}, which only partially resembles conventional waterfilling. We are now ready to state the following theorem.
\begin{thm}
The iterative ADRMP algorithm that updates sequentially the power of each user $k\in\mathcal{K}$ according to $\eqref{eq:linProb}$ converges to a Nash equilibrium that is also a (local) maximizer for the global rate of the system $R(\mathbf{p})$.
\end{thm}
\begin{IEEEproof}
To prove this theorem, we need to show first that the solution of ADRMP in  \eqref{eq:linProb} is a better response for user $k$. Let us assume that the current  strategy for user $k$ is $\mathbf{y}_{k}$ and let $\mathbf{x}_{k}\in\mathcal{P}_{k}$ be the power distribution obtained as solution of \eqref{eq:linProb} with $\mathbf{p}_{k}(0)=\mathbf{y}_{k}$, the following inequality holds
\begin{equation}
R\left(\mathbf{y}_k,\mathbf{p}_{-k}\right) \leq \tilde{R}\left(\mathbf{x}_k,\mathbf{p}_{-k};\mathbf{y}_k\right)\leq R\left(\mathbf{x}_k,\mathbf{p}_{-k}\right)
\label{property}
\end{equation}
The first inequality follows from $R\left(\mathbf{y}_k,\mathbf{p}_{-k}\right)= \tilde{R}\left(\mathbf{y}_k,\mathbf{p}_{-k};\mathbf{y}_k\right)\leq \tilde{R}\left(\mathbf{x}_k,\mathbf{p}_{-k};\mathbf{y}_k\right)$. The second inequality descends directly from the fact that in \eqref{eq:linProb} we approximate part b) of \eqref{eq:rateMaxDist}, which is a convex function, with its tangent in $\mathbf{y}_k$ and thus by definition of convexity is $\tilde{R}\left(\mathbf{x}_k,\mathbf{p}_{-k};\mathbf{y}_k\right)\leq R\left(\mathbf{x}_k,\mathbf{p}_{-k}\right)$.
Since the power allocation game $\mathcal{G}\big(\mathcal{K},\{\tilde{\mathcal{P}}_{k}\},R(\mathbf{p}_{k},\mathbf{p}_{-k})\big)$ is an exact potential game, the set $\tilde{\mathcal{P}}$ is continuous and compact and the iterative strategy based on \eqref{eq:linProb} that sequentially updates the users' power profiles is a better response dynamic, the game converges  to a pure NE, which is also a maximizer of the potential function, i.e. the global rate $R(\mathbf{p})$.
\end{IEEEproof}
Given a fixed scheduling order $\pi(\mathcal{K})$, Algorithm \ref{AlgI} illustrates the iterative procedure to allocate the power among the $K$ D2D tx-rx couples: the algorithm is iterated until the difference between the overall rate computed in two successive iterations does not exceed a certain threshold $\epsilon$, whose value depend on the system designer. At each iteration only one user updates its power, while all other users do not change their strategies. In particular, in row 5 the new power allocation for user $\ell$ at iteration $j+1$ is computed solving problem \eqref{eq:linProb} with $\mbf{p}_{\ell}(0)=\mbf{p}_{\ell}^{(j)}$.

\begin{algorithm}[t]
	\small \caption{Iterative ADRMP} \label{AlgI}
	\begin{algorithmic}[1]
	\Statex \emph{Initialization}
	\State  $j\gets 0$
	\State Set $\mathbf{p}^{(0)}\gets\mbf{p}^{(ini)} $,  $\Delta\gets\epsilon$
	\State Compute $R(\mathbf{p}^{(0)})$
	
	\Statex $j+1$  recursion given the power vector $\mathbf{p}^{(j)}$
	\While { $\Delta\ge\epsilon$}
		\For {$\ell \in \pi(\mathcal{K})$}
				\State{$\mathbf{p}_{-\ell}^{(j)}\gets[\mathbf{p}_{1}^{(j+1)}, \mathbf{p}_{2}^{(j+1)},\dots,\mathbf{p}_{\ell-1}^{(j+1)},\mathbf{p}_{\ell+1}^{(j)},\dots,\mathbf{p}_{K}^{(j)}]$}
				\State $\mathbf{p}_{\ell}^{(j+1)}\gets\arg\max\limits_{\mathbf{x}\in\mathcal{P}_{\ell}}\tilde{R}\left(\mathbf{x},\mathbf{p}_{-\ell}^{(j)};\mathbf{p}_{\ell}^{(j)}\right)$ subject to $\mathbf{1}^{T}\mathbf{x} \le P_{\ell}$
		\EndFor
		\State Compute $R(\mathbf{p}^{(j+1)})$ 	
		\State  $\Delta\gets \max\limits_{k\in\mathcal{K}}\left(R\left(\mathbf{p}_{k}^{(j+1)},\mathbf{p}_{-k}^{(j+1)}\right)-R\left(\mathbf{p}_{k}^{(j)},\mathbf{p}_{-k}^{(j)}\right)\right)$
		\State $j\gets j+1$
	\EndWhile

	\end{algorithmic}
\end{algorithm}
%

\subsection{A Multi-start approach to the solution of \eqref{eq:rateMaxJoint}}\label{sec:multistart}
Since there might be more than one maximum for the JRMP, it is not straightforward to assess the `optimality' of the local maximum found with the iterative technique described in  Algorithm \ref{AlgI}. In cases like this,  a  practical means of addressing a global optimization  problem might be to run a local optimization routine several times starting it from many different points and to select the best solution among those found. This approach, sometimes termed \emph{controlled randomization}  \cite{glover1986future} or \emph{multi-start} \cite{marti2013multi}, in principle does not guarantee that a global maximum is found, but increased confidence can be gained by using a large number of starting points accurately chosen. Among the infinite points $\mathbf{p}\in\tilde{\mathcal{P}}$ that can be employed to start the iterative ADRMP algorithm, there is a finite set determined by the user scheduling order, which directly translates into a priority ordering. If, for example, is $\mbf{p}^{(ini)}=\mbf{0}$, at the beginning of iteration $j=1$ the first user allocates its power in absence of any interference  and is free to select the resources that are best for her, the second user sees already a certain amount of interference and so on until the last user, whose  power allocation is very much influenced by the allocations priorly made by the other users. This applies also for iteration $j>1$ when the users' choices are  in any case influenced by the power allocations made at iteration $j-1$. In practice, each different  scheduling order and starting power vector $\mbf{p}^{(ini)}$ might  determine a new local maximum. Since the number of different scheduling  configurations is  finite, it is possible to run the iterative power allocation algorithm for all the  user scheduling configurations to produce a set of different local maxima and to chose the user ordering that achieves the maximum rate within this set.

Although not feasible in practice due to its complexity and the amount of control information required, this multi-start approach is a heuristic method to find the global maximum or a very near approximation of it and it can be employed as a benchmark for the performance of the iterative ADRMP algorithm applied to the D2D dedicated mode.

\section{D2D Reuse Mode: Allocation Problem with Interference Constraints}

In reuse operation mode D2D communications take place underlaying the primary cellular network. In particular, we consider the case where the D2D network shares the available spectrum with the uplink transmissions of the UEs. To allow the coexistence of the two transmissions, we follow an approach derived from cognitive radio theory \cite{Zhang2008} and we constrain the transmit power at the D2D nodes  so that the received interference at the eNB is below a given predetermined threshold on each subcarrier.

Let  $Q_{n}$ be the threshold value for the interference caused  by the D2D nodes at the eNB on subcarrier $n$, the JRMP with interference constraints (JRMPIC) power allocation  can be formulated as a different version of  \eqref{eq:rateMaxJoint} with a new set of constraints
\begin{align}
\max\limits_{\mathbf{p}\in\mathcal{P}} &\sum \limits_{k \in\mathcal{K}} \sum\limits_{n \in \mathcal{N}}\log_{2}\left(1+\frac{p_{k,n}}{i_{k,n}}\right) \nonumber \\
&\st \label{eq:linProb_newC}\\
&\sum_{n\in\mathcal{N}}p_{k,n}\leq P_{k} \quad k\in\mathcal{K} \nonumber\\
& \sum_{k\in\mathcal{K}} A^{n}_{k,0}p_{k,n} \leq Q_{n} \quad n\in\mathcal{N}  \nonumber
\end{align}
where $A^{n}_{k,0}$ is the squared absolute value of the channel gain on subcarrier $n$ between the D2D transmit user of couple $k$ and the eNB. By setting the values of $Q_{n}$ the eNB can implicitly select the specific D2D mode: very high values of $Q_{n}$ make problem \eqref{eq:linProb_newC} practically equivalent to problem \eqref{eq:rateMaxJoint}, while $Q_{n}=0$ prevents any D2D node from using subcarrier $n$.

The JRMPIC problem is formulated to control the amount of interference that D2D communications cause to the primary cellular network. On the other hand, the interference of primary network transmissions on the D2D secondary network, which it is not controllable by the D2D network, is encompassed into the noise terms $\sigma_{k,n}^{2}$.

Although  some of the constraints of \eqref{eq:linProb_newC}  are formulated as the global sum of the interference at the eNB, we will show that,  provided that the eNB broadcasts  some information back to the D2D network, JRMPIC can still be solved by employing the distributed game theoretic approach discussed in the previous Section.

\subsection{An Upper Bound for JRMPIC}\label{Subgradient_UB}

The solution of JRMPIC  can be upper-bounded in the Lagrangian dual domain, where the constraints on maximum tolerated interference at the eNB are relaxed and a different Lagrange multiplier is associated to each constraint. The  Lagrangian of problem \eqref{eq:linProb_newC} can be written as
\begin{equation}
\label{eq:Lagrangian}
\mathcal{L}(\mathbf{p},\boldsymbol{\nu})= \sum\limits_{k \in \mathcal{K}}\sum\limits_{n \in \mathcal{N}}\log_{2}\left(1+\frac{p_{k,n}}{i_{k,n}}\right) + \sum\limits_{n \in \mathcal{N}}\nu_{n} \left( Q_{n}-\sum_{q\in\mathcal{K}}  A^{n}_{q,0}p_{q,n} \right)
\end{equation}
where $\mathbf{p}$ belongs to the set of feasible power vectors $\tilde{\mathcal{P}}=\left\{\mathbf{p} \in \mathcal{P} \vert \sum_{n\in\mathcal{N}}p_{k,n}\leq P_{k}, \forall k \in\mathcal{K}\right\}$ and $\boldsymbol{\nu}=[\nu_{1},\nu_{2},\dots,\nu_{N}]$ is the vector of Lagrange multipliers associated to the set of constraints on maximum tolerated interference at the eNB. The Lagrange dual function is computed by maximizing the Lagrangian with respect to the primal variable $\mathbf{p}$ as
\begin{equation}
\label{eq:linProb_newC_rel}
g(\boldsymbol{\nu})=\max\limits_{\mathbf{p}\in\tilde{\mathcal{P}}}\mathcal{L}(\mathbf{p},\boldsymbol{\nu}).
\end{equation}
\par
The Lagrangian in \eqref{eq:Lagrangian} is not convex in $\mbf{p}$ and, just as  in \eqref{eq:rateMaxJoint}, a local maximum of $\mathcal{L}(\mathbf{p},\boldsymbol{\nu})$ with respect to $\mbf{p}$ can be found  by letting each user $k\in\mathcal{K}$ solve iteratively a distributed problem in $\mathbf{p}_{k}$. Neglecting the terms not dependent on $\mathbf{p}_{k}$, the distributed maximization problem for user $k$ can be formulated as
\begin{equation}
\label{eq:linProb_newC_rel1}
\max\limits_{\mathbf{p}_{k}\in\tilde{\mathcal{P}_{k}}}  \sum\limits_{n \in \mathcal{N}}\log_{2}\left(1+\frac{p_{k,n}}{i_{k,n}}\right)+\sum\limits_{n \in \mathcal{N}}\alpha_{k,n}p_{k,n} - \sum\limits_{n \in \mathcal{N}}\nu_{n}A^{n}_{k,0}p_{k,n},
\end{equation}
where $\alpha_{k,n}$ is computed as in \eqref{eq:Gradient}. Replacing  $\alpha_{k,n}$ with the term $\alpha'_{k,n}=\alpha_{k,n}-\nu_{n}A^{n}_{k,0}$, the optimization in \eqref{eq:linProb_newC_rel1}  is formally identical to \eqref{eq:linProb} and it can be solved in the same manner, adopting the multi-start approach discussed in Sect. \ref{sec:multistart} in combination with the  iterative ADRMP algorithm. The assumption that problem \eqref{eq:linProb_newC_rel} has been solved optimally is of great importance because  it guarantees the convexity of the dual function $g(\boldsymbol{\nu})$ \cite{Boyd2004}.
\par
Being able to compute $g(\boldsymbol{\nu})$ in \eqref{eq:linProb_newC_rel}, one can formulate the Lagrange dual problem, whose solution is a bound for \eqref{eq:linProb_newC}, as
\begin{align}
& \min\limits_{\bs{\nu}}  g(\bs{\nu})\nonumber \\
& \st \label{eq:linProb_newC_rel0}\\
&\bs{\nu}\succeq 0 \nonumber
\end{align}
Problem \eqref{eq:linProb_newC_rel0}  is convex and a standard approach for finding its solution is to follow an iterative strategy, such as the \emph{ellipsoid method} illustrated in Appendix B, which recursively updates the vector of Lagrange variables until convergence \cite{MoPN13}. In this case there are two nested iterative algorithms: an outer one that iterates on the vector of Lagrange multipliers $\bs{\nu}$ to solve \eqref{eq:linProb_newC_rel0} and an inner one that, given a value of $\bs{\nu}$, solves  \eqref{eq:linProb_newC_rel} yielding $g(\bs{\nu})$ and the corresponding optimal power vector $\mbf{p}(\bs{\nu})$.  A key element for the outer iterative procedure is the availability of the gradient or, if $g(\bs{\nu})$ is not differentiable with respect to $\bs{\nu}$ as is our case, at least of a subgradient of the Lagrangian dual function. Adapting to our problem Proposition 1 of \cite{Yu06}, one can show that  the vector $\mathbf{d}(\bs{\nu})=[d_{1}(\bs{\nu}),d_{2}(\bs{\nu}),\dots,d_{N}(\bs{\nu})]^{T}$, whose $n$th element is computed as
\begin{equation}
\label{eq:subGrad}
d_{n}(\bs{\nu})  =Q_{n}-\sum_{q\in\mathcal{K}} A^{n}_{q,0}p_{q,n}(\bs{\nu})
\end{equation}
is a subgradient for $g(\bs{\nu})$. Therefore, given the ellipsoid $\mathcal{E}\left(\mathbf{A}^{(s)},\boldsymbol{\nu}^{(s)}\right)$ as in \eqref{eq:ellipDef}, the  $s$th iteration of the algorithm  designed to solve \eqref{eq:linProb_newC_rel0}  can be summarized as
\begin{enumerate}
\item Plug $\bs{\nu}^{(s)}$ in \eqref{eq:linProb_newC_rel} and compute $g(\bs{\nu}^{(s)})$;
\item Employ the power vector $\mathbf{p}(\bs{\nu}^{(s)}) \in\tilde{\mathcal{P}}$, solution of the maximization in \eqref{eq:linProb_newC_rel}, to compute the subgradient $\mathbf{d}\left(\bs{\nu}^{(s+1)}\right)$ as in \eqref{eq:subGrad};
\item  Find the ellipsoid $\mathcal{E}\left(\mathbf{A}^{(s+1)},\boldsymbol{\nu}^{(s+1)}\right)$ by means of equations \eqref{eq:EM1}-\eqref{eq:EM3} in Appendix B.
\end{enumerate}
The value of the Lagrange dual function $g(\bs{\nu}^{\ast})$ at convergence is an upper bound of the solution of JRMPIC, which can be employed to validate heuristic algorithms designed to solve \eqref{eq:linProb_newC} sub-optimally.

\subsection{A Practical and Distributed approach for JRMPIC}\label{Subgradient_dem}
In a practical D2D scenario, the multi-start strategy is unviable because it is too complex and requires far too many iterations and too much coordination among terminals. Employing the iterative ADRMP algorithm without the multi-start strategy to solve \eqref{eq:linProb_newC_rel1} leads to finding  a \emph{local} maximum of the Lagrangian.
Nevertheless, one can draw inspiration from the algorithm presented in the previous section and pursue  a heuristic  approach based on the  relaxation of the original problem with respect to the interference constraints and employ two nested  iterative algorithms  to find a sub-optimal feasible solution of JRMPIC. In the following, in continuity with the notation used in the previous sections we will indicate with the apex $s$ the iteration index relative to the outer loop designed to find the vector of multipliers  $\boldsymbol{\nu}$ and with the apex $j$ the iteration index relative to the inner power control loop.

The main difference with the algorithm introduced in the previous subsection is that, since we are not able to solve exactly problem \eqref{eq:linProb_newC_rel}, we propose a heuristic strategy where the outer iterative algorithm is based on the auxiliary function  $\tilde{g}^{(s)}(\bs{\nu}^{(s)})=\mathcal{L}(\tilde{\mbf{p}}^{(s)},\bs{\nu}^{(s)})$, where the power vector $\tilde{\mbf{p}}^{(s)}=[\tilde{\mbf{p}}^{(s)}_{1},\tilde{\mbf{p}}_{2}^{(s)},\dots,\tilde{\mbf{p}}_{K}^{(s)}]^{T}$ does not necessarily achieve the global maximum since it is just a local maximizer of $\mathcal{L}(\mbf{p},\bs{\nu}^{(s)})$, obtained by iteratively solving \eqref{eq:linProb_newC_rel1}. In particular, since the value of $\tilde{\mbf{p}}^{(s)}$ depends on: $\bs{\nu}^{(s)}$, the starting power vector $\mbf{p}^{(ini,s)}$ and the scheduling order $\pi$, we assume that $\pi$ is fixed and that at each iteration $s$, the starting power vector  is the solution of the previous local maximization, i.e. $\mbf{p}^{(ini,s)}=\tilde{\mbf{p}}^{(s-1)}$. This particular choice is motivated by the need of algorithm speed and stability.

Under these hypothesis, we can now introduce a new lemma about the properties of $\tilde{g}^{(s)}(\bs{\nu})$.
\begin{lem}
Let $\tilde{\mbf{p}}^{(s)}(\bs{\nu})$ the power vector at iteration $s$, when the vector of Lagrange multipliers is $\bs{\nu}$. The following inequality holds for $\tilde{g}^{(s+1)}(\bs{\mu})$ with any $\bs{\mu}\succeq\mbf{0}$
\begin{equation}
\label{eq:subGradDef}
\tilde{g}^{(s+1)}(\bs{\mu})\ge\tilde{g}^{(s)}(\bs{\nu})+\tilde{\mbf{d}}^{T}(\bs{\nu})\left(\bs{\mu}-\bs{\nu}\right).
\end{equation}
where  $\tilde{\mbf{d}}(\bs{\nu})$ is the $N$-dimensional vector whose entries are $\tilde{d}_{n}(\bs{\nu}) =Q_{n}-\sum_{q\in\mathcal{K}} A^{n}_{q,0}\tilde{p}^{(s)}_{q,n}(\bs{\nu})$ ($n=1,\dots,N$).
\end{lem}
\begin{IEEEproof}  By definition it is
\begin{equation}\label{Lagmod}
\tilde{g}^{(s)}(\bs{\nu})=\mathcal{L}(\tilde{\mbf{p}}^{(s)}(\bs{\nu}),\bs{\nu}) = R(\tilde{\mbf{p}}^{(s)}(\bs{\nu})) +\sum\limits_{n \in \mathcal{N}}\nu_{n} \left( Q_{n}-\sum_{q\in\mathcal{K}}  A^{n}_{q,0}\tilde{p}^{(s)}_{q,n}(\bs{\nu})\right)
\end{equation}
Keeping in mind that $\mbf{p}^{(ini,s+1)}=\tilde{\mbf{p}}^{(s)}(\bs{\nu})$, i.e. at step $s+1$ the iterative algorithm is initialized with the power vector $\tilde{\mbf{p}}^{(s)}(\bs{\nu})$, one can write regardless of the value of $\bs{\mu}$
\begin{align}
\label{eq:subGradDem}
\tilde{g}^{(s+1)}(\bs{\mu})&=\mathcal{L}(\tilde{\mbf{p}}^{(s+1)}(\bs{\mu}),\bs{\mu}) \nonumber\\
&\ge\mathcal{L}(\tilde{\mbf{p}}^{(s)}(\bs{\nu}),\bs{\mu})=R(\tilde{\mbf{p}}^{(s)}(\bs{\nu})) + \sum\limits_{n \in \mathcal{N}}\mu_{n} \left( Q_{n}-\sum_{q\in\mathcal{K}}  A^{n}_{q,0}\tilde{p}^{(s)}_{q,n}(\bs{\nu}) \right)\\
&=R(\tilde{\mbf{p}}^{(s)}(\bs{\nu})) + \sum\limits_{n \in \mathcal{N}}\nu_{n} \left( Q_{n}-\sum_{q\in\mathcal{K}}  A^{n}_{q,0}\tilde{p}^{(s)}_{q,n}(\bs{\nu}) \right)+\sum\limits_{n \in \mathcal{N}}(\mu_{n}-\nu_{n}) \left( Q_{n}-\sum_{q\in\mathcal{K}}  A^{n}_{q,0}\tilde{p}^{(s)}_{q,n}(\bs{\nu})\right) \nonumber\\
&=\tilde{g}^{(s)}(\bs{\nu})+\tilde{\mbf{d}}^{T}(\bs{\nu})(\bs{\mu}-\bs{\nu}). \nonumber
\end{align}
The first inequality in \eqref{eq:subGradDem} is due to the better response property of the distributed algorithm \eqref{eq:linProb_newC_rel1}: each new solution is larger than the previous one and, since $\tilde{\mbf{p}}^{(s)}(\bs{\nu})$ is by definition the starting value and $\tilde{\mbf{p}}^{(s+1)}(\bs{\mu})$ is the power vector at convergence, then it is $\mathcal{L}(\tilde{\mbf{p}}^{(s+1)}(\bs{\mu}),\bs{\mu})\ge\mathcal{L}(\tilde{\mbf{p}}^{(s)}(\bs{\nu}),\bs{\mu})$.
\end{IEEEproof}
The inequality \eqref{eq:subGradDef} closely resembles a subgradient for the  the auxiliary function  $\tilde{g}^{(s)}(\bs{\nu})$ and accordingly we apply a subgradient update rule to the vector of Lagrangian multipliers, ie
\begin{equation}
\bs{\nu}^{(s+1)}= \left[\bs{\nu}^{(s)} - \gamma\tilde{\mbf{d}}(\bs{\nu}^{(s)})\right]^+
\end{equation}
where $\gamma$ is a sufficiently small step size.

Algorithm \ref{AlgII} illustrates the machinery of the outer loop of the heuristic designed for solving JRMPIC, which we will indicate with the acronym iterative ADRMPIC. The algorithm is iterated until the maximum difference in power per user does not exceed a given arbitrarily small value $\epsilon$.
\begin{algorithm}[h]
	\small \caption{Iterative ADRMPIC} \label{AlgII}
	\begin{algorithmic}[1]
	\Statex \emph{Initialization}
	\State  $s\gets 0$, $\bs{\nu}^{(0)}\gets\mbf{0}$, $\tilde{\mbf{p}}^{(0)}\gets\mbf{p}^{(ini)}$, $\Delta\gets\epsilon$
	\State $\tilde{d}_{n}(\bs{\nu})^{(0)} \gets Q_{n}-\sum_{q\in\mathcal{K}} A^{n}_{q,0}\tilde{p}^{(0)}_{q,n}\quad \forall n\in\mathcal{N}$			
	\Statex $s+1$  recursion given the multiplier vector $\bs{\nu}^{(s)}$
	\While { $\Delta\ge\epsilon$}
		\State $\bs{\nu}^{(s+1)}\gets \left(\bs{\nu}^{(s)} - \gamma \tilde{\mbf{d}}(\bs{\nu}^{(s)})\right)^{+}$
		\State Compute  $\tilde{\mbf{p}}^{(s+1)}$ and $\tilde{g}^{(s+1)}(\bs{\nu}^{(s+1)})$ by employing Algorithm \ref{AlgI} to solve 	\eqref{eq:linProb_newC_rel} with $\mbf{p}^{(ini,s+1)}=\tilde{\mbf{p}}^{(s)}$
		\State $\tilde{d}_{n}(\bs{\nu}^{(s+1)}) \gets Q_{n}-\sum_{q\in\mathcal{K}} A^{n}_{q,0}\tilde{p}^{(s+1)}_{q,n}\quad\forall n\in\mathcal{N}$	
		\State $\Delta= \max\limits_{k\in\mathcal{K}}\|\tilde{\mbf{p}}_{k}^{(s+1)}-\tilde{\mbf{p}}_{k}^{(s)}\|_{2}$		
		\State $s\gets s+1$
	\EndWhile

	\end{algorithmic}
\end{algorithm}
%

\subsection{Extension of JRMPIC to a multi cell scenario}

The JRMPIC can be easily formulated in a multi-cell scenario and its solution, except for a few details, does not change substantially with respect to the single-cell scenario. To elaborate, let us refer to a general cellular setting and introduce the set $\mathcal{B} = \left\{0,\ldots,B-1 \right\}$ of the eNBs in the system and denote by $Q_{b,n}$ the maximum interference  tolerated at the $b$th eNB on subcarrier $n$.  For notational convenience we still indicate by $\mathcal{K}$ the whole set of D2D couples, without specifying to which cell each node belongs. In this case, the JRMPIC power allocation problem can be formulated as
\begin{align}
\max\limits_{\mathbf{p}\in\mathcal{P}} &\sum \limits_{k \in\mathcal{K}} \sum\limits_{n \in \mathcal{N}}\log_{2}\left(1+\frac{p_{k,n}}{i_{k,n}}\right) \nonumber \\
&\st \label{eq:linProb_newCMulti}\\
&\sum_{n\in\mathcal{N}}p_{k,n}\leq P_{k} \quad k\in\mathcal{K} \nonumber\\
& \sum_{k\in\mathcal{K}} A^{n}_{k,b}p_{k,n} \leq Q_{b,n} \quad n\in\mathcal{N}, b\in\mathcal{B} \nonumber
\end{align}
Problem \eqref{eq:linProb_newCMulti} is almost identical to Problem \eqref{eq:linProb_newC} and can be solved with the algorithms devised for the single-cell scenario, with the difference that, in this case, the vector of Lagrange multipliers $\bs{\nu}=[\nu_{1,1},\nu_{1,2},\dots,\nu_{N,B}]$ accounts for the $NB$ interference constraints, $N$ for each cell.
%
%

\section{On the Implementation of Distributed Power Allocation}

The proposed iterative ADRMP and ADRMPIC algorithms are naturally amenable to a distributed implementation, where all involved terminals act independently. Indeed, better response dynamics guarantee convergence with a fully random scheduling, without any coordination with the other D2D users, neither in the same cell nor in adjacent cells.
Nevertheless, to solve allocation  problem \eqref{eq:linProb} the $k$th D2D node requires the knowledge of the term $\alpha_{k,n}$, which accounts for how its power allocation  impacts on the performance of the other terminals. \par
Distributing the messages $\alpha_{k,n}$ to the D2D terminals may require 
the help of the eNB, which can collect the messages from all D2D nodes under its coverage and broadcast them to all active D2D transmitters. Moreover, when needed, eNBs in adjacent cells can  exchange the messages among each other by using a proper inter-cell communication interface, e.g., the X2-Interface in LTE \cite{LTE-A1}.
To sum up, as in other distributed power control schemes \cite{Papandriopoulos2009}, the proposed power allocation algorithms  present a strong predisposition towards distributed implementation but they rely on wide message passing between all involved nodes, and as such may suffer from some overhead.

In a TDD scenario, an alternative strategy that does not require any eNB involvement consists in letting each D2D node  broadcast a sounding signal using a proper in-band control channel, which does not interfere with direct communications. Wideband sounding reference signals (SRS), which span all available subcarriers, are already envisaged in LTE \cite{LTE-A1} for estimating the uplink channel of connected terminals across the scheduling bandwidth. It is also possible to exploit the SRS for accomplishing control tasks among D2D terminals, as for example proposed in \cite{D2D-discovery2}.

In detail, from \eqref{eq:Gradient} we can factorize $\alpha_{k,n}=\sum_{\ell \in \mathcal{K}\setminus k}G_{k,\ell}^{n}\delta_{\ell,n}$, where  $\delta_{\ell,n}$ defined as:
\begin{equation}
\label{eq:Gradient2v}
\delta_{k,n} = \frac{G_{k,k}^{n}p_{k,n}}{\ln2\left(\sum\limits_{j\in \mathcal{K}\setminus \{k\}}G_{j,k}^{n}p_{j,n}+\sigma_{k,n}^{2}\right)\left(\sum\limits_{j\in \mathcal{K}\setminus \{k\}}G_{j,k}^{n}p_{j,n}+G_{k,k}^{n}p_{k,n}+\sigma_{k,n}^{2}\right)}
\end{equation}
can be measured at the $\ell$th D2D receiver  and needs to be signaled to all other terminals.   Hence, by letting  the $\ell$th receive node transmit over subcarrier $n$ a sounding  signal with power $\delta_{\ell,n}p_{0}$ and the $b$th eNB a sounding signal with power $\nu_{b,n}p_{0}$, where $p_0$ is  a fixed power factor known at each terminal, the power measured over channel $n$ at the $k$th transmitter
is $\sum_{\ell \in \mathcal{K}}G_{k,\ell}^{n}\delta_{\ell,n}p_{0}-\sum_{b\in \mathcal{B}}A_{k,b}^{n}\nu_{b,n}p_{0} = \alpha'_{k,n}+G_{k,k}^{n}\delta_{k,n}p_{0}$. An estimate of $\alpha'_{k,n}$ can be obtained by subtracting the term $G_{k,k}^{n}\delta_{k,n}p_{0}$, which is known at  the $k$th D2D transmitter  by exploiting the dedicated control channel with the $k$th receiver, and explicit message passing can be completely avoided.

\section{Numerical results}\label{sec: num res}

In this section we present the numerical results of the proposed
algorithms. We have considered an hexagonal cell of radius $R=500$ m. Channel attenuation is due to path loss,
proportional to the distance between transmitters and receivers, shadowing and fading. The path
loss exponent is $\alpha=4$, while the shadowing is assumed log-normally distributed with standard deviation $\sigma = 8$ dB. We consider a
population of data users with very limited mobility so that the
channel coherence time can be assumed very long. The propagation channel is
frequency-selective Rayleigh with independent fading coefficients on each subcarrier. The variance of the additive zero-mean Gaussian noise, which includes the interference from the infrastructured network, is set to $10^{-13}$ W, the same for all receivers and for all subcarriers, i.e., $\sigma^{2}_{k,n} = \sigma^{2} = 10^{-13}$.
The number of subcarriers is set to $N = 8$ and the maximum power constraint $P_{k}$ is assumed to be the same for all D2D couples, and equal to $P_{max} = 0.25$ W, when not indicated otherwise. The power mask $P_{k,n}$ for user $k$ on subcarrier $n$ is determined by the maximum allowed interference at the serving eNB, i.e., $P_{k,n} = Q_{b(k),n}/G^{n}_{k,b(k)}$, where $b(k)$ is the index of the eNB which serves the $k$-th D2D couple. Eventually, the number of D2D couples is set to $K = 8 \times B$, i.e., we consider $8$ D2D couples per cell. Hence, at each simulation instance the D2D couples are deployed randomly in the cell, with a tx-rx distance uniformly distributed in the interval $[0,D_{max}]$, with $D_{max} = 100$ m.
\par
In the D2D overlay scenario we compare  the performance  of the iterative ADRMP (IADRMP) scheme presented in Algorithm \ref{AlgI} with the performance of the  classical iterative waterfilling (IWF) algorithm \cite{YuRhee2004}, the SCALE algorithm proposed in \cite{Papandriopoulos2009}  and the near-optimal multi-start solution, denoted in the following by IADRMP-MS. The IADRMP algorithm is implemented employing  a fixed scheduling order $\pi$ for all simulations, and choosing the  initial power allocation $\mbf{p}^{(ini)}$ as  the power vector obtained by each user solving problem \eqref{eq:rateMaxJoint} without considering the interference terms generated by other users, i.e., by setting $G_{j,k}^{n}p_{j,n} = 0$, $\forall$ $k,n$.
The IWF scheme can be  derived from IADRMP by setting $\alpha_{k,n} = 0$ and, since its  convergence is not always guaranteed, its performance is evaluated terminating the simulation after a sufficiently high number of iterations. As for the SCALE algorithm, it makes use of successive convex approximations so that the original problem can be decomposed into a sequence of convex subproblems, which are solved iteratively until convergence. The SCALE algorithm requires  that all nodes exchange messages among them: each node upon receiving its messages simultaneously updates its transmitting power, i.e., implementing a network-wide parallel update rule. For this reason, the distributed implementation  of SCALE in a wireless network is more complex with respect to the proposed IADRMP scheme, where all nodes update their powers independently.
\begin{figure}[ht]
\begin{center}
\includegraphics[width=8 cm]{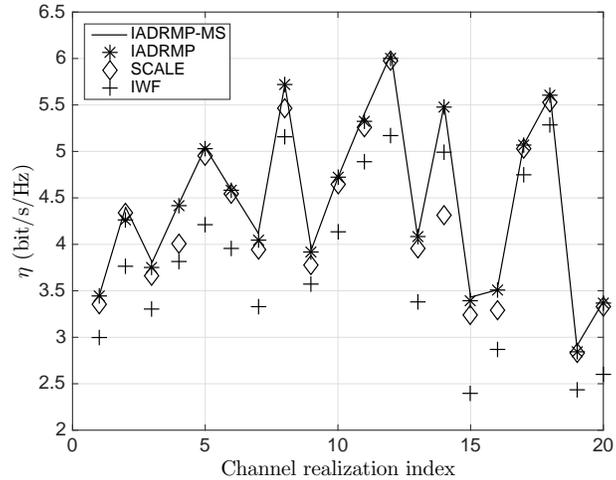}
\caption{
Spectral efficiency for IADRMP-MS, IADRMP, SCALE and IWF allocation schemes, obtained for 20 different channel realizations, in the case of $B = 1$.
}
\label{F1}
\end{center}
\end{figure}
\begin{figure}[ht]
\begin{center}
\includegraphics[width=8 cm]{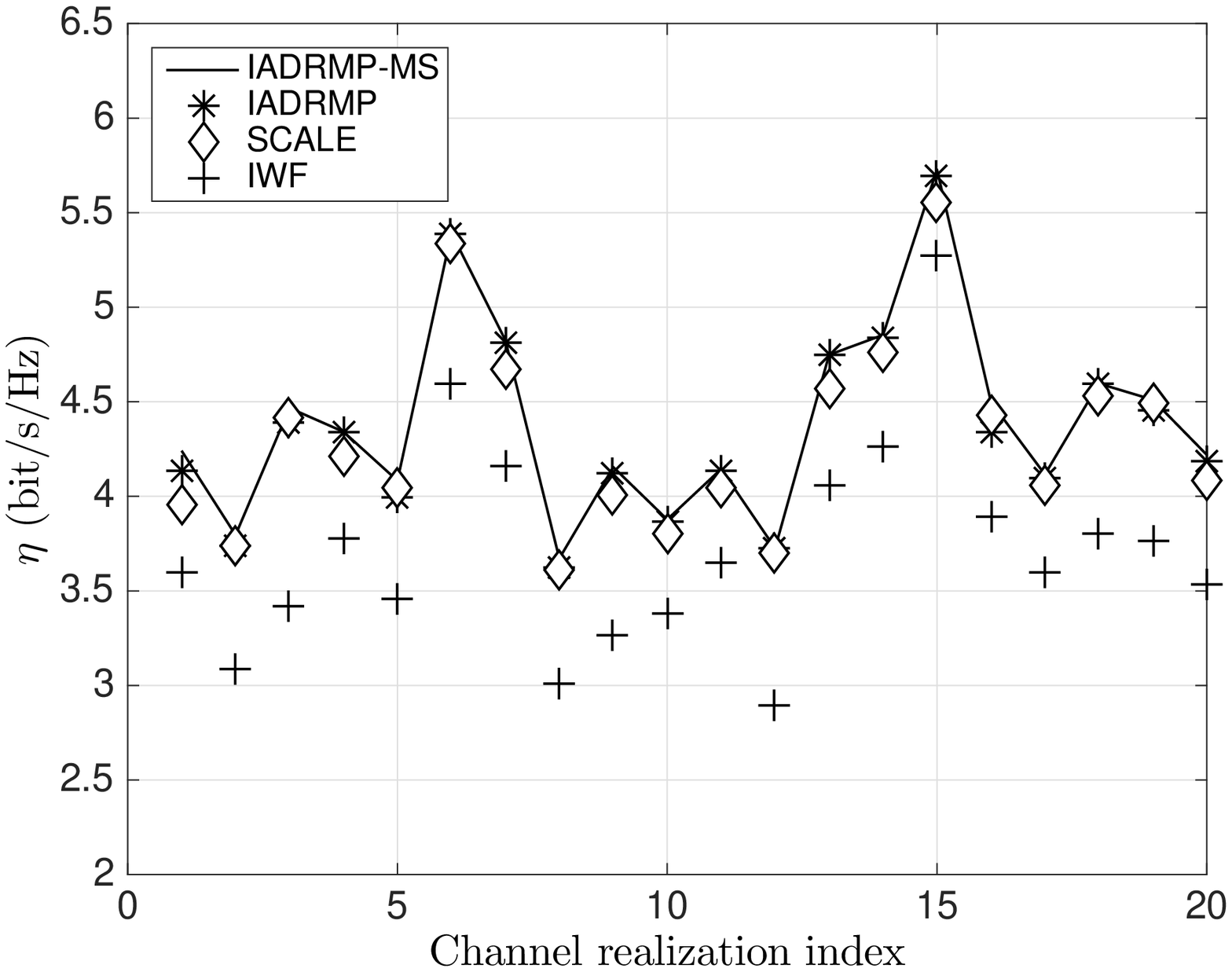}
\caption{
Spectral efficiency for IADRMP-MS, IADRMP, SCALE and IWF allocation schemes, obtained for 20 different channel realizations, in the case of $B = 3$.
}
\label{F2}
\end{center}
\end{figure}
\begin{figure}[ht]
\begin{center}
\includegraphics[width=8 cm]{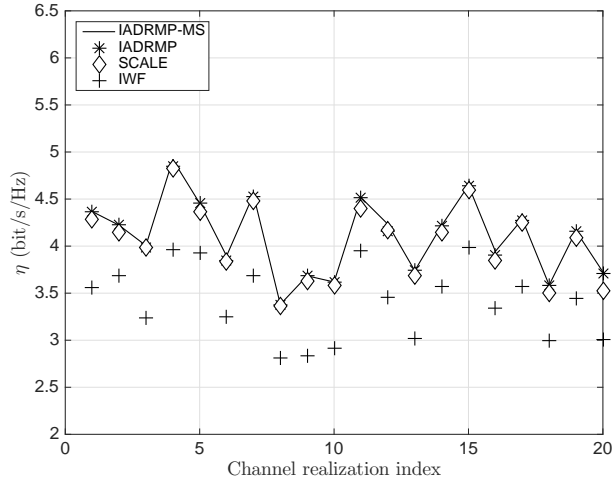}
\caption{
Spectral efficiency for IADRMP-MS, IADRMP, SCALE and IWF allocation schemes, obtained for 20 different channel realizations, in the case of $B = 7$.
}
\label{F3}
\end{center}
\end{figure}
\par
Figs. \ref{F1}-\ref{F3} plot the achieved spectral efficiency $\eta$ per cell measured as the sum rate per cell  over the bandwidth for the various allocation schemes obtained for 20 different channel realizations in the case of a system with  $B = 1$, $B = 3$ and $B = 7$ cells, respectively.  We observe that in most of the considered channel realizations, IADRMP  outperforms the SCALE algorithm and its performance is very close to that of the ADRMP-MS scheme. On the other hand, IWF performs significantly worse than the other schemes in all the considered cases.
\begin{table}[ht]
\centering
\renewcommand{\arraystretch}{1.1}
\begin{tabular}{c| c| c| c| c| c|}
                       \hline
\multicolumn{1}{|c|}{ } & IADRMP-MS & IADRMP   & SCALE  & IWF \\ \hline
\multicolumn{1}{|c|}{$B$ = 1}  &285.26 &283.33  &273.41  &246.43\\ \hline
\multicolumn{1}{|c|}{$B$ = 3}  &844.72 &837.49  &825.67  &714.92\\ \hline
\multicolumn{1}{|c|}{$B$ = 7}  &1840.06 &1833.02  &1807.49  &1527.7 \\ \hline
\end{tabular}
\caption{Aggregated throughput averaged over 100 different channel realizations for IADRMP-MS, IADRMP, SCALE and IWF.}
\label{tab.results}
\end{table}
\par
More extensive results, obtained by averaging the aggregated throughput over 100 different channel realizations, are shown in Table \ref{tab.results}. In general, as the number of eNBs increases the achievable spectral efficiency is reduced but all algorithms show that they are able to efficiently deal with both inter- and intra-cell interference and the difference between the optimal IARDMP-MS and IADRMP tend to vanish.
\par
\begin{figure}[ht]
\begin{center}
\includegraphics[width=8 cm]{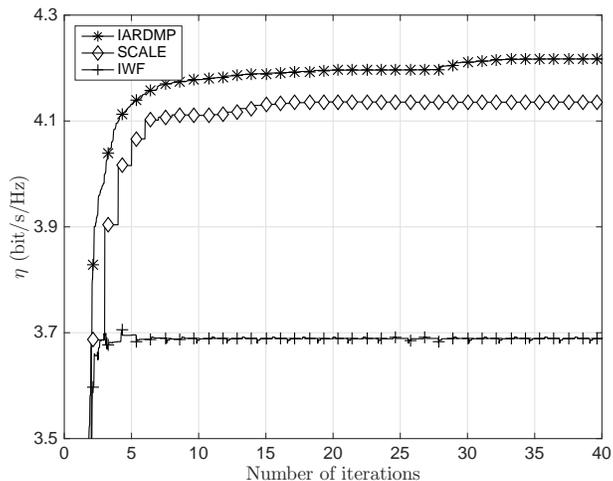}
\caption{Convergence behavior of IADRMP, SCALE and IWF in the case $B = 7$, for a single simulation realization.}
\label{F4}
\end{center}
\end{figure}

Fig. \ref{F4} reports the convergence behavior of IADRMP, SCALE and IWF. To this aim, we show the aggregated throughput versus the number of iterations in the case $B = 7$ for a single channel snapshot. More precisely, we count any cicle $j$ in Algorithm 1 as one iteration. The convergence speed of IADRMP and SCALE is similar, whereas, as expected, IWF keeps on fluctuating without achieving convergence. Similar results are obtained considering different realizations and are omitted here for the sake of conciseness.
\par
To analyze the performance of the proposed algorithms in the D2D underlay scenario, we compare the  iterative ADRMPIC (IADRMPIC) scheme presented in Algorithm \ref{AlgII} with its  upper bound derived in Section \ref{Subgradient_UB}, referred to as IADRMPIC-UB. Note that, since SCALE and IWF schemes are not designed to cope with global interference constraints, they are not considered in the results for the D2D reuse mode.
The maximum allowed interference at the eNB is set to the power of the AWGN noise  on each subcarrier of each cell, i.e., $Q_{b,n} = \sigma^{2}$, $\forall b\in\mathcal{B},n\in\mathcal{N}$.
The initial power allocation $\mbf{p}^{(ini)}$  for the IADRMPIC scheme is set as discussed for the IADRMP case, while for the IADRMPIC-UB  we iteratively run the multi-start scheme to optimally solve \eqref{eq:linProb_newC_rel} for each $\boldsymbol{\nu}$, where $\boldsymbol{\nu}$ are updated according to the ellipsoid method. This  task is computationally very expensive, particularly when the dimension of the problem is high. For this reason, we limit the evaluation of the IADRMPIC-UB performance to the single cell case.
\par
\begin{figure}[ht]
\begin{center}
\includegraphics[width=8 cm]{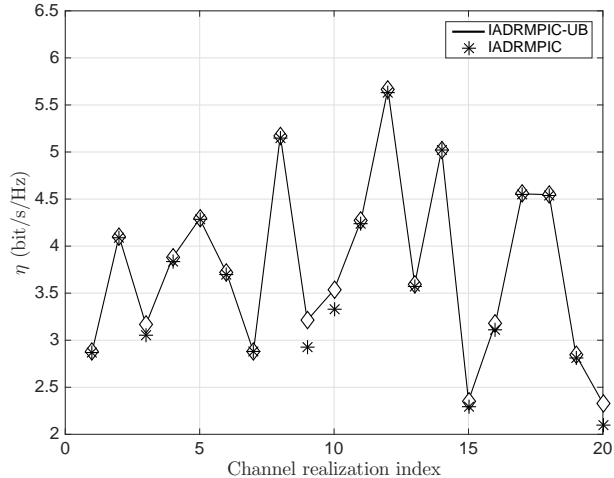}
\caption{
Spectral efficiency for IADRMPIC-UB, and IADRMPIC obtained for 20 different channel realizations, in the case of $B = 1$ and for $Q_{max} = \sigma^{2}$.
}
\label{F5}
\end{center}
\end{figure}

Fig. \ref{F5} plots the spectral efficiency results for IADRMPIC-UB and IADRMPIC, obtained for 20 different channel realizations in the case $B = 1$. In most of the considered channel realizations  IADRMPIC achieves performance very close to bound. As a matter of fact, the average aggregated throughput obtained over the considered channel realizations is 235.9 for IADRMPIC and 240.3 for IADRMPIC-UB, i.e., the IADRMPIC performs worse by nearly $2\%$ with respect to the upper bound.
\par
\begin{figure}[ht]
\begin{center}
\includegraphics[width=8cm]{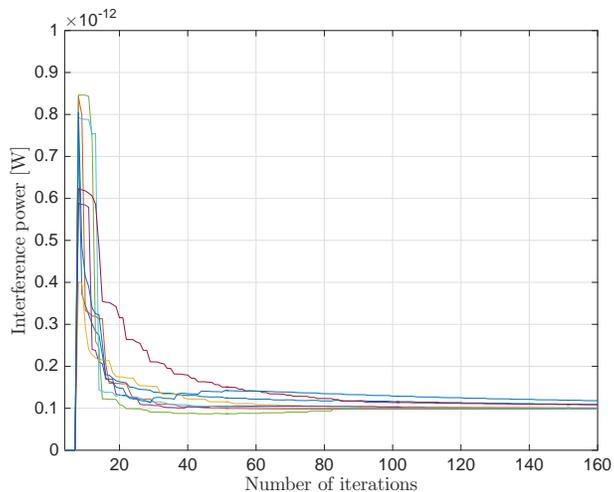}
\caption{
Interference experienced at the eNB on all the available subcarriers versus the number of iterations, for a single simulation realization in the case of $B = 7$ and for $Q_{b,n} = \sigma^{2}$.
}
\label{F6}
\end{center}
\end{figure}

Fig. \ref{F6} shows the convergence behavior of IADRMPIC for the case $B = 7$ for a single simulation realization by plotting  the  interference experienced at the eNB in the central cell on all the subcarriers versus the number of iterations. The results show the merit of the heuristic approach proposed: in a reasonably small number of iterations the interference power on each subcarrier is close to the target $Q_{b,n}=\sigma^{2}$. Similar results are obtained considering different realizations and are omitted here for brevity.
\par
\begin{figure}[ht]
\begin{center}
\includegraphics[width=8cm]{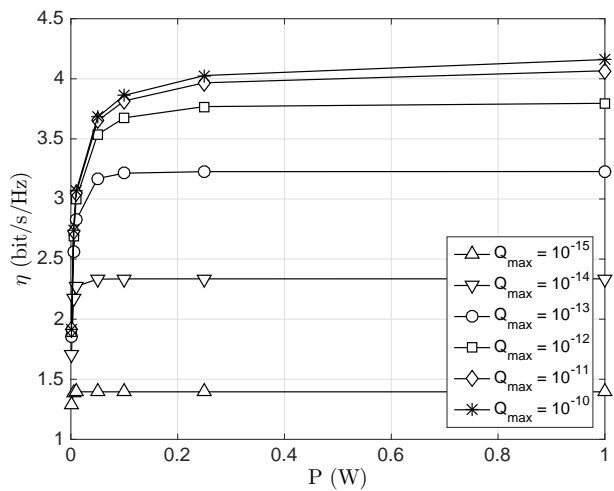}
\caption{
Spectral efficiency for IADRMPIC as a function of $P$ for different interference constraints $Q_{max}$ in the case B = 7.
}
\label{F8}
\end{center}
\end{figure}

In Fig. \ref{F8} we report the spectral efficiency for IADRMPIC as a function of the power constraints. In this case we set $B = 7$, $P_{k}=P$ $\forall k\in\mathcal{K}$, $Q_{b,n}=Q_{max}$ $\forall n\in\mathcal{N}$  and assign  different values to the maximum interference constraint $Q_{max}$. When $Q_{max}=10^{-15}$, the performance of the D2D nodes are dominated by the interference constraints so that already at low power levels any power increase does not result in any efficiency increment. Gradually, as more interference is tolerated at eNB also the spectral efficiency of the D2D nodes grows proportionally. For $Q_{max}>10^{-11}$ the interference constraints are not binding in most of the cases and the performance mainly depend on the available power. The case of $Q_{max}=10^{-10}$ is roughly equivalent to the optimization in the overlay scenario where no interference constraints are present at all.
\par
\begin{figure}[ht]
\begin{center}
\includegraphics[width=16cm]{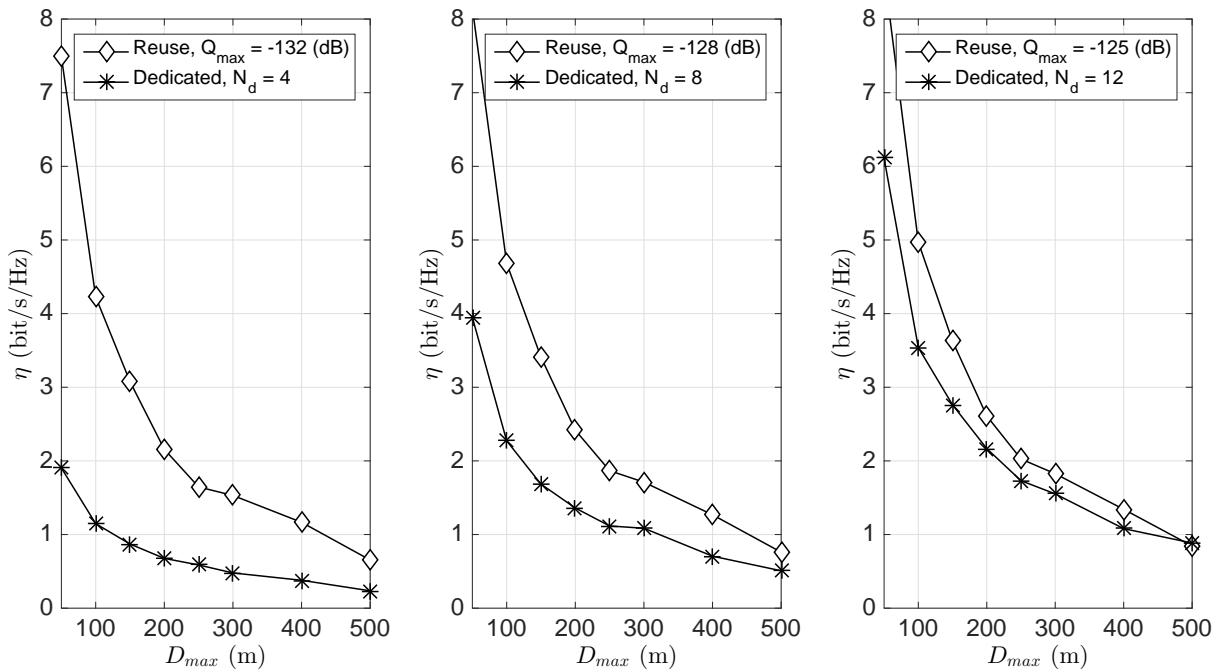}
\caption{
Spectral efficiency for dedicated and reuse mode versus the maximum distance $D_{max}$ between each D2D pair in the case B = 3.
}
\label{F8b}
\end{center}
\end{figure}

Fig. \ref{F8b} compares the performance of the dedicated  and reuse modes by plotting the spectral efficiency $\eta$, achieved by the IADRMP and IADRMPIC schemes, respectively, versus the maximum distance $D_{max}$ between each D2D pair. In particular, the  three different scenarios
  (a), (b) and (c)  represent the cases in which $12.5\%$, $25\%$ and $50\%$ of the available resources are  dedicated to D2D communications. The spectral efficiency is computed as the average bit rate of the D2D connections normalized by the the whole system bandwidth, i.e., the normalization factor is the same in the three scenarios. In this case we slightly change the simulation settings with respect to the previous figures and consider a cellular environment with $N = 24$ available subcarriers and $B=3$ cells, each serving 8 UEs and 4 D2D pairs.
Regarding the infrastructured network, we assume that the available subcarriers are   assigned uniformly to the UEs and that  each UE allocates its power employing the waterfilling algorithm with a power budget $P_{max} = 1$ W and a fixed interference-plus-noise term for each subcarrier, given by $Q_{max} + \sigma^{2}$.

For the dedicated case the three different scenarios of Fig. \ref{F8b} are obtained by setting  $N_d$, the number of subcarriers which are reserved to D2D communications, to $N_d = 4$, $8$, and $12$, respectively.
To perform a fair comparison between the two D2D modes, the value of the parameter $Q_{max}$ for the reuse  mode  is  set so that the effect of the D2D interference  on the UEs' throughput   is completely compensated by the availability of a larger number of subcarriers and the total throughput of cellular UEs is exactly the same as that obtained in the dedicated mode. Hence, the higher $N_d$, the worse the performance of cellular UEs (on account of the minor bandwidth), and, accordingly, the higher the tolerated interference $Q_{max}$, whose value increases from $Q_{max}=-132$ in scenario (a) to $Q_{max}=-125$ in scenario (c).


In all scenarios $\eta$ decreases with the increase of $D_{max}$, but such an effect is more evident in the reuse case due to the stringent constraints on the interference at the eNB. It is also worth noting that, as expected, any increment of  $N_d$ causes an increase of  $\eta$ in the dedicated case since $N_d$ is a measure of the actual bandwidth available for D2D connections and $\eta$ is computed by normalizing the total D2D rate by the total system bandwidth. 
In line with this reasoning, the higher $Q_{max}$ the better is the reuse mode performance since there is a higher level of D2D interference tolerated at the eNB. In both modes, any improvement achieved by the D2D network is obtained at the expense of the performance of the UEs in the infrastructured network.
 The curves in Fig. \ref{F8b}  show that, with these simulation settings, despite the the fact that the reuse gain diminishes with the increase of $D_{max}$, the more flexible reuse mode, implemented with the proposed IADRMPIC allocation scheme, always outperforms the dedicated mode being able to  more efficiently exploit  the available radio resources.

\section{Conclusions}\label{sec: conclusion}

We have presented a distributed resource allocation framework for D2D communication considering both dedicated and reuse mode. As for the dedicated mode, due to the NP-hardness of original resource allocation problem, we have invoked some important game theoretic results about potential games to find a distributed iterative solution for the rate maximization problem which provably converges to a local maximum. For D2D reuse mode the allocation problem is formulated with an additional requirement for each subcarrier so that the total interference generated at the base station by the D2D nodes does not exceed a given threshold. Accordingly, after finding the optimal solution, which is too complex for practical implementation, we propose a heuristic algorithm, which builds on the power allocation algorithm devised for the dedicated D2D mode to find a feasible solution. Hence, we have discussed about possible practical implementations of the proposed allocation schemes. In particular, we have proposed an approach based on the use of a broadcast sounding signal, so that the required information to perform power allocation can be gathered from interference measurements, without requiring neither message passing nor additional channel gain estimations. Numerical simulations, carried out for several different user scenarios, show that the proposed methods, which converge to one of the local maxima of the objective function, exhibit performance close to the maximum achievable optimum and outperform other schemes presented in the literature. Moreover, comparisons between the proposed allocation schemes in a typical cellular scenario, where cellular and D2D UEs coexist in the same area, assess the superiority of the reuse mode, thus proving the effectiveness of the proposed allocation schemes in exploiting the available radio resources.

\section*{Appendix A: Solution ADRMP \eqref{eq:linProb}} \label{sec:AppA}
\renewcommand{\thesection}{\Alph{section}}
\renewcommand{\theequation}{A.\arabic{equation}}

\setcounter{equation}{0}
In this appendix we derive the solution of the linearized power allocation problem \eqref{eq:linProb}, which, for ease of readability, we rewrite here
\begin{align}
\max\limits_{\mathbf{p}_{k}\in\mathcal{P}_{k}} & \sum\limits_{n \in \mathcal{N}}\log_{2}\left(1+\frac{p_{k,n}}{i_{k,n}}\right)+\sum\limits_{n \in \mathcal{N}}\alpha_{k,n}p_{k,n} \nonumber \\
& \st \label{eq:linProbApp}\\
& \sum_{n\in\mathcal{N}}p_{k,n}\leq P_{k} \nonumber
\end{align}
where $i_{k,n}=\frac{\sum\limits_{j\in \mathcal{K}\setminus k}G_{j,k}^{n}p_{j,n}+\sigma_{k,n}^{2}}{G_{k,k}^{n}}$ is the noise plus interference term normalized to the $k$th user gain.
The problem is convex with differentiable objective and constraint function and, hence, any points that satisfy the KKT conditions are primal and dual optimal and have zero duality gap. The problem's Lagrangian is
\begin{equation}
\mathcal{L}(\mathbf{p}_{k},\mu)=\sum\limits_{n \in \mathcal{N}}\log_{2}\left(1+\frac{p_{k,n}}{i_{k,n}}\right)+\sum\limits_{n \in \mathcal{N}}\left(\alpha_{k,n}-\mu\right)p_{k,n}+\mu P_{k}
\end{equation}
where $\mu$ is the dual variable associate to the user's total power constraint.
Accordingly, we find that the optimum allocation $\mathbf{p}^{\ast}_{k}\in\mathcal{P}_{k}$ must satisfy the following KKT conditions:
\begin{align}
\sum_{n\in\mathcal{N}}p_{k,n}-P_{k}&\leq 0 \\
\mu &\ge 0\\
\mu\left(\sum_{n\in\mathcal{N}}p_{k,n}-P_{k}\right) &= 0 \label{eq:CS}  \\
\frac{1}{\log(2) \left(i_{k,n}+p_{k,n}\right)}+\alpha_{k,n}-\mu &= 0
\label{KKT}
\end{align}
As a consequence of the \emph{complementary slackness} condition  \eqref{eq:CS} when $\mu=0$ it is $\sum_{n\in\mathcal{N}}p_{k,n}< P_{k}$ and when $\mu >0$ it is $\sum_{n\in\mathcal{N}}p_{k,n}= P_{k}$. Thus, unlike conventional waterfilling, user $k$ might not need to use all the available power $P_{k}$.  By elaborating \eqref{KKT} and assuming that $\sum_{n\in\mathcal{N}}p^{\ast}_{k,n}< P_{k}$, the optimal solution is
\begin{equation}
p^{\ast}_{k,n} = \left[-\frac{1}{\log(2)}\frac{1}{\alpha_{k,n}}-i_{k,n}\right]^{P_{k,n}}_{0}
\label{eq:optimalPow_mu0}
\end{equation}
where
\begin{equation}
\left[ x \right]^{A}_{0}=\left\{
\begin{array}{ll}
0 & \textrm{$x<0$}\\
x & \textrm{$0 \leq x \leq A$}\\
A & \textrm{$A < x$}
\end{array}
\right.
\end{equation}
In case the power distribution found in \eqref{eq:optimalPow_mu0} exceeds the power limit $P_{k}$, we have to assume that $\mu>0$ and power is found as
\begin{equation}
p^{\ast}_{k,n} = \left[\frac{1}{\log(2)}\frac{1}{\mu-\alpha_{k,n}}-i_{k,n}\right]^{P_{k,n}}_{0}
\label{eq:optimalPow_muNot0}
\end{equation}
where the value of $\mu$ is such that the power constraint $P_{k}$ is met.
\section*{Appendix B: The Ellipsoid Method} \label{sec:AppB}
\renewcommand{\thesection}{\Alph{section}}
\renewcommand{\theequation}{B.\arabic{equation}}
\setcounter{equation}{0}

The ellipsoid method is an iterative technique that  starts with the ellipsoid
\begin{equation}
\label{eq:ellipDef}
\mathcal{E}\left(\mathbf{A}^{(0)},\boldsymbol{\nu}^{(0)}\right)= \left \{z \in \mathbb{R}^{N} : (z - \boldsymbol{\nu}^{(0)})^T \mathbf{A}^{(0)} (z-\boldsymbol{\nu}^{(0)}) \leq 1 \right \}
\end{equation}
centered in  $\boldsymbol{\nu}^{(0)}$ and with a shape defined by the symmetric and  positive definite matrix $\mathbf{A}^{(0)}$. By choosing appropriate values for $\mathbf{A}^{(0)}$ and $\boldsymbol{\nu}^{(0)}$, the ellipsoid $\mathcal{E}\left(\mathbf{A}^{(0)},\boldsymbol{\mu}^{(0)}\right) $ contains the solution $\boldsymbol{\nu}^{*}$ of  problem \eqref{eq:linProb_newC_rel} and, by construction, at each iteration the algorithm finds a  new ellipsoid that still contains the solution $\boldsymbol{\nu}^{*}$ but with a smaller volume.
Hence, given an arbitrary small volume $\epsilon$,  after a certain number of iterations  the ellipsoid's volume will be smaller than $\epsilon$. Thus, we can  choose an adequate  value of $\epsilon$, such that the centre of the ellipsoid practically coincides with $\boldsymbol{\nu}^{*}$.

Given the subgradient vector $\mathbf{d}^{(s)}$, the update rule for the ellipsoid algorithm for iteration $s$ is \cite{Boyd2004}:
\begin{equation}
\tilde{\mathbf{d}}^{(s)}=\frac{\mathbf{d}^{(s)}}{\sqrt{\mathbf{d}^{(s)T}\mathbf{A}^{(s)-1}\mathbf{d}^{(s)}}}
\label{eq:EM1}
\end{equation}
\begin{equation}
\boldsymbol{\mu}^{(s+1)}=\boldsymbol{\mu}^{(s)}-\frac{1}{N+1}\mathbf{A}^{(s)-1}\tilde{\mathbf{d}}^{(s)}
\end{equation}
\begin{equation}
\mathbf{A}^{(s+1)}=\frac{N^{2}}{N^{2}-1}\left(\mathbf{A}^{(s)-1}-\frac{2}{N+1}\mathbf{A}^{(s)-1}\tilde{\mathbf{d}}^{(s)}\tilde{\mathbf{d}}^{(s)T}\mathbf{A}^{(s)-1}\right).
\label{eq:EM3}
\end{equation}

\bibliographystyle{IEEEtran}
\bibliography{D2Dref,Biblio_paper_long}

\end{document}